# Electronic band structure of Two-Dimensional WS$_2$/Graphene van der Waals Heterostructures

Hugo Henck[1], Zeineb Ben Aziza[1], Debora Pierucci[2], Feriel Laourine[1], Francesco Reale[3] and Pawel Palczynski[3], Julien Chaste[1], Mathieu G. Silly[4], François Bertran[4], Patrick Le Fevre[4], Emmanuel Lhuillier[5], Taro Wakamura[6], Cecilia Mattevi[3], Julien E. Rault[4], Matteo Calandra[5] and Abdelkarim Ouerghi[1,*]

[1]Centre de Nanosciences et de Nanotechnologies, CNRS, Univ. Paris-Sud, Université Paris-Saclay, C2N – Marcoussis, 91460 Marcoussis, France
[2]CELLS - ALBA Synchrotron Radiation Facility, Carrer de la Llum 2-26, 08290 Cerdanyola del Valles, Barcelona, Spain
[3]Imperial College London Department of Materials, Exhibition road London SW7 2AZ, UK
[4]Synchrotron-SOLEIL, Saint-Aubin, BP48, F91192 Gif sur Yvette Cedex, France
[5]Sorbonne Universités, UPMC Univ. Paris 06, CNRS-UMR 7588, Institut des NanoSciences de Paris, 4 place Jussieu, 75005 Paris, France
[6]Laboratoire de Physique des Solides, CNRS, Université Paris-Sud, Université Paris Saclay, 91405 Orsay Cedex, France

*Corresponding author, E-mail: **abdelkarim.ouerghi@c2n.upsaclay.fr**

**Abstract:**

Combining single-layer two-dimensional semiconducting transition metal dichalcogenides (TMDs) with graphene layer in van der Waals heterostructures offers an intriguing means of controlling the electronic properties through these heterostructures. Here, we report the electronic and structural properties of transferred single layer WS$_2$ on epitaxial graphene using micro-Raman spectroscopy, angle-resolved photoemission spectroscopy measurements (ARPES) and Density Functional Theory (DFT) calculations. The results show good electronic properties as well as well-defined band arising from the strong splitting of the single layer WS$_2$ valence band at K points, with a maximum splitting of 0.44 eV. By comparing our DFT results with local and hybrid functionals, we find the top valence band of the experimental heterostructure is close to the calculations for suspended single layer WS$_2$. . Our results provide an important reference for future studies of electronic properties of WS$_2$ and its applications in valleytronic devices.

**Keywords:** WS$_2$ – Spectroscopy – Electronic properties– Spin splitting – Interlayer interaction.

## I. INTRODUCTION

Two-dimensional (2D) materials, such as graphene, h-BN, phosphorene, and transition metal dichalcogenides (TMDs), are an ideal platform for studying exciting physical properties not attainable in their bulk counterparts [1]. They often exhibit a versatile electronic structure controllable by thickness [2], surface chemical adsorption, and strain [3]. Of particular interest are the family of TMDs of formula $MX_2$ (M = Mo, W; X=S, Se). They are semiconductor with unique properties in the 2D limit such as indirect to direct band gap transition, large exciton binding energy, well-defined valley degrees of freedom, and spin splitting of the valence band [4]. Many efforts have been made to harvest these properties into practical optoelectronic, spintronic, and valleytronic devices [5].

In addition to homogenous 2D materials, van der Waals (vdW) heterostructures have recently emerged as a novel class of materials, in which different 2D atomic planes are vertically stacked to give rise to distinctive properties and exhibit new structural, chemical, and electronic phenomena [1,6]. These heterostructures, in contrast with traditional heterostructures, can be designed and assembled by stacking individual 2D layers without lattice parameter constraints. The weak interlayer coupling in vdW heterostructures offers the possibility of combining the intrinsic electronic properties of each individual 2D layers in devices. In particular, $WS_2$/graphene vdW heterostructures are remarkable since they gather the high carrier mobility and broadband absorption of graphene [7], as well as the direct bandgap and extremely strong light-matter interactions of single layer $WS_2$. The combination of these unusual characteristics led to potential applications in field-effect transistor devices [8] and energy harvesting materials [9].

$MX_2$ single layer, the elementary unit to form ultrathin films by weak stacking, features a novel spin-valley coupled band structure. At the corners of the first Brillouin zone (BZ), the valence (conduction) band has two inequivalent valleys. Owing to the broken inversion symmetry in single layers, the strong spin-orbit coupling (SOC) from the d-orbitals of the metal atoms results in a valence band spin splitting at K points, with a magnitude as large as 400 meV in $WS_2$ [10,11]. The spin-splitting has opposite signs at the K and K' valleys as they are time reversal of each other at the valence band. This spin-valley coupling forms the basis for manipulation of spin and valley degrees of freedom in these 2D semiconductors.

Angle resolved photoemission spectroscopy (ARPES) is one of the most suitable tools to investigate the electronic structure of 2D materials with energy and momentum resolution as well as surface sensitivity [12], [13], [14], [15]. There have been a number of ARPES studies either on bulk $MX_2$ samples or few-layer samples prepared by various methods such as exfoliation, chemical vapor deposition (CVD) [16,17], and epitaxial growth [18,19]. They all observe sizable splitting in the VB at the K-point, with a size of ~150 meV for $MoS_2$ and ~400 meV for $WS_2$ as tungsten induces larger SOC [20]. Furthermore, uniform and large-area synthesis of single layer $WS_2$ is an important subject toward applications of 2D materials in various electronic devices [21].

## I. METHODS:

Single layer graphene was produced via a two-step process beginning with a starting substrate of 4H-SiC(0001). Prior to graphitization, the substrate was hydrogen etched (100% $H_2$) at 1550 °C to produce well-ordered atomic terraces of SiC. Subsequently, the SiC sample was heated to 1000 °C and then further heated to 1550 °C in an Ar atmosphere [22–24]. $WS_2/SiO_2$ samples were grown by chemical vapor deposition (CVD) in a 1" quartz tube furnace. The growth substrate was placed in the center of the furnace and heated to 800 °C. A 25 mg sulfur pellet was placed on a piece of silicon and positioned upstream in the furnace such that its temperature was approximately 150 °C. Carrier gas (500 sccm $N_2$) was used to bring sulfur vapor into the furnace for a 30 min growth period.

The ARPES measurements were conducted at the CASSIOPEE beamline of Synchrotron SOLEIL (Saint-Aubin, France). We used linearly polarized photons of 50 eV and a hemispherical electron analyzer with vertical slits to allow band mapping. The total angle and energy resolutions were 0.25° and 10 meV. All ARPES experiments were done at low temperature (8 K). XPS experiments were carried out on the TEMPO beamline of Synchrotron SOLEIL (Saint-Aubin, France) at room temperature. The photon source was a HU80 Apple II undulator set to deliver linearly polarized light. The photon energy was selected using a high-resolution plane grating monochromator with a resolving power E/ΔE that can reach 15,000 on the whole energy range (45 - 1500 eV). During the XPS measurements, the photoelectrons were detected at 0° from the sample surface normal $\vec{n}$ and at 46° from the polarization vector $\vec{E}$.

## I. RESULTS AND DISCUSSIONS

In this work, TMD/graphene heterostructures were made from WS$_2$ grown by CVD on SiO$_2$/Si substrates that were then transferred onto graphene/SiC [25] (more details about the transfer procedure are given in the supplementary information Section I) [26]. A schematic of WS$_2$ crystal structure is presented in Figure 1(a) where the lattice constants are indicated. The CVD growth of WS$_2$ on SiO$_2$ results in characteristic single-crystal domains shaped as well-defined equilateral triangles [27]. These films can be nondestructively transferred to various kinds of substrates as desired. With respect to SiO$_2$ and similar substrates, graphene layer has favorable qualities such as atomic flatness, and homogeneous charge distribution. This should enable direct investigation of the adjacent TMD's intrinsic electronic structure and many-body effects. Graphene layer is often used as a substrate for TMD heterostructures with high device performance [28]. Single layered WS$_2$ flakes were then easily identified by their optical contrast with respect to the graphene substrate (Figure 1(b)) and confirmed by micro-Raman spectroscopy (Figure 1 (c) and (d)). The optical image in Figure 1(b) shows large (lateral size about 50 μm) triangular flakes of WS$_2$ on the graphene layer. The graphene underlayer used in this study was obtained by annealing 4H-SiC(0001) (see Methods). After the transfer of WS$_2$ onto graphene, an annealing process at T = 300 °C for 60 minutes in UHV was used to further clean the surface and interface of the WS$_2$/graphene heterostructure. To investigate the structural properties of the WS$_2$ flakes micro-Raman and photoluminescence (PL) spectroscopy were used (Figure 1 (c) and (d) and supplementary information section II and III). Several Raman spectra of WS$_2$ on graphene, collected in different flake positions, are shown in Figure 1(c). These Raman spectra are obtained with 532 nm excitation, which is in resonance with the B exciton absorption peak [29,30] . Then, beside the first order modes at the Brillouin zone (BZ) center (Γ), the in-plane phonon mode $E_{2g}^1$ at 356 cm$^{-1}$ and the out-of-plane phonon mode A$_{1g}$ at 418 cm$^{-1}$ [31,32], the Raman spectra present a series of overtone and combination peaks. These different contributions are clearly separated using a multi-peak Lorentzian fit (blue lines in Figure 1 (c) top). In particular, the Raman feature around 350 cm$^{-1}$ is the convolution of several components: the $E_{2g}^1$ ($\Gamma$), the 2 LA (M) mode at 351.7 cm$^{-1}$, which is a second-order Raman mode due to LA phonons at the M point of the BZ zone, and the $E_{2g}^1$ ($M$) mode [33]. Moreover, Raman peaks at 323.7 cm$^{-1}$ and 297.1 cm$^{-1}$ are combinations modes, which are attributed to $2LA\ (M) - E_{2g}^2\ (\Gamma)$ and the $2LA\ (M) - 2E_{2g}^2\ (\Gamma)$ modes, respectively. Raman mapping was conducted to investigate the uniformity within the single flake. The $2LA\ (M)$, $E_{2g}^1\ (\Gamma)$ and the A$_{1g}$ ($\Gamma$) modes intensity and position mapping are shows in Figure 1(d) and S2. These Raman modes peak positions and intensity are uniform within the single crystal, indicating that the electronic properties of WS$_2$ are uniform on the graphene substrate. From the frequency difference between the A$_{1g}$ ($\Gamma$) and the 2LA(M), the thickness of the WS$_2$ flakes can be determined [34–36]. An average distance of $65.9 \pm 0.1$ cm$^{-1}$ is obtained, as shown in figure S4. This value is consistent with monolayer WS$_2$ as already shown by previous studies [37]. For the following experiments, the single layer coverage was estimated to be around 20% of the total area of the sample.

The high quality of the WS$_2$ transferred on graphene layer allows investigating the electronic structure by X-ray photoelectron spectroscopy (XPS) and ARPES. Both the feasible large-area synthesis and the reliable film transfer process can promise that WS$_2$ ultrathin films will pave a route to many applications of 2D materials and vdW heterostructure. XPS is used to determine the chemical composition and stoichiometry of the WS$_2$ films transferred on the graphene. The C-1s XPS spectrum of WS$_2$/graphene heterostructure, collected at hν=340 eV, is shown in Figure 2(a). The C-1s spectrum showed three components at 283.9, 284.7, and 285.1 eV in binding energy (BE). These components correspond to the SiC bulk (labeled SiC), the graphene layer (labeled G), and the interface layer (labeled I), respectively. The G peak was fitted by a sum of a Gaussian function convoluted with a Doniach-Sunjic line shape with an asymmetry factor α of 0.09 and a FWHM of 0.4 eV. The low value of the FWHM indicates that only one core level peak was present and thus the carbon atoms had a unique chemical environment. Moreover, no structure appeared at ~286.7 eV in the C-1s XPS spectrum, usually attributed to the contamination and/or oxidation, this means even if the samples were annealed at 300 °C, the WS$_2$/graphene layers were very inert and did not show any contamination after WS$_2$ transfer process.

The W-4f and S-2p XPS peaks deconvolution shown in Figure 2(b) and (c), present the standard WS$_2$ stoichiometry. The WS$_2$ is well fitted by a doublet peaks at BE of 33.4 eV, 35.6 eV, corresponding to W-4f$_{7/2}$, W-4f$_{5/2}$ core energy levels (4f$_{5/2}$: 4f$_{7/2}$ ratio of 0.75) [38], respectively (Figure 2(b)). The sulfur S-2p peak (Figure 2(c)) consists of a single doublet (S 2p$_{3/2}$ at a BE = 163 eV and a 2p$_{1/2}$:2p$_{3/2}$ ratio of 0.5 and spin-orbit splitting of 1.19 eV) corresponding to S-W bonding, and confirms the presence of WS$_2$. The absence of the oxygen content [39,40] in these samples is a result of high quality of the interface in this hybrid hetero-structure, particularly in light of the Raman characteristics that are comparable to high quality WS$_2$ (see above and supplementary information section II and ref [41]).

The electronic structure was also probed using ARPES measurements. The WS$_2$ flakes are relatively well spaced on the graphene substrate, ensuring, thanks to the small spot size (50×50 μm$^2$), the mapping of a single flake. Figure 2(d) shows the photoelectron intensity as a function of energy and k-momentum around the normal emission. This is a good measure of the band structure of WS$_2$ around Γ. The zero of binding energy (*i.e.* the Fermi level) was determined by fitting the leading edge of the graphene layer at the same photon energies and under the same experimental conditions. Beside the typical dispersion of the π bands of graphene around -6 to -8 eV (most intense band), a new set of bands is visible at the Γ point of the BZ, which is the signature of WS$_2$ valence band. In the spectra, the most distinct features include the valence band maximum (VBM) at Γ. The full width at half maximum (FWHM) of WS$_2$ band branches is about 80 meV as shown by Energy distribution curves (EDC) in Figure S7) extracted from the ARPES map of Figure 2(d) at the Γ point. The sharpness of the different bands can be attributed to the high quality of the transferred flake. Figure 2(e) shows the measured band structure corresponding to the graphene underlayer. The single and robust Dirac cone confirms that the graphene monolayer at the heterostructure preserves the Dirac linear dispersion and the massless relativistic character of the graphene carriers close to the Fermi level. The Dirac point (E$_D$) is located at 0.40 eV below the Fermi level as in the case of pristine graphene on SiC [12]. Then, the n-type doping, close to $9 \times 10^{12}$ cm$^{-2}$, of pristine graphene was not modified by the formation of the heterostructure. This confirms, as shown also by the work function measurement (see supplementary information

Figure S8), that there was no significant charge transfer between the 2D materials. However, if we look more in detail at the momentum distribution curves (MDCs) shown in figure 2 (f) a less dispersive behavior of the peaks are present around $E_D$. Then, the presence of gap-like feature at the K point (already present in the pristine graphene) cannot be completely excluded [42]. However, a more detail study of this feature goes beyond the scope of this paper. From a linear fitting of the MDCs a Fermi velocity $v_F \sim 1.05 \times 10^6$ m/s was also determined.

It is obvious that also the sharp and intense structure of π-bands confirms the high structural quality of $WS_2$ / graphene heterostructure. Differently from our previous work on $MoS_2$/graphene layer and from Diaz et al. [12,43] no signature of interlayer hybridization and mini-gaps opening [44] is present on the π-band of graphene. The absence of this superperiodicity effects probably is related to the mismatch angle between the $WS_2$ flake and the graphene underlayer. In the following discussions regarding the ARPES data, we will focus mostly on the split of the valence band feature at the K-point of the $WS_2$ (Figures 3). In particular, in order to gain more insight on the electronic structure of $WS_2$, we performed first principles electronic structure calculations on an isolated $WS_2$ monolayer using the QUANTUM ESPRESSO code [45,46] and PBE [47] and HSE06 [48] hybrid functionals within the adaptive compressed exchange [49] implementation. More technical details are given in the supplemental materials (section IV). Figure 3(a) shows the measured band structure around the K valley. The top of the valence band at the K point is mostly formed by planar $d_{xy}$ and $d_{x2-y2}$ orbital of tungsten, while at the Γ point the band is mostly composed by W $d_z$ orbitals and S $p_z$ orbitals.

The observation of a single valence band at Γ with a higher binding energy than at K also excludes the contribution from bilayer or trilayer $WS_2$: in bilayer TMD systems, the valence band near Γ shows a bonding/anti-bonding splitting that would be observable by ARPES as two bands, in contrast to what is seen here [15]. The maximum of the valence band is located at the K point (-1.8 eV, which is 0.3 eV higher than at Γ point). Since the band gap of single $WS_2$ is 2.27 eV [50], these measurements indicate that our sample is heavily electron doped, which is consistent with previous reports. [51] The electronic structure in the -2 eV to -3 eV binding energy range is globally in good agreement with our electronic structure calculations (Figures 3(b) and (c)), although the energy separation between the top of the valence band at K and Γ is slightly underestimated in PBE (Figure 3 (b)). In both theory and experiments the valence band maximum at the Γ point is located at higher binding energies than that at the K-point, the energy separation between the two being 0.20 eV in experiments (Figure 3(b) and (c)), 0.17 eV in PBE and 0.20 eV in HSE Figure 3 (c). The measured spin-orbit splitting at K is about 440 meV. Our theoretical calculations for the isolated $WS_2$ monolayer lead to splittings of 440 meV (PBE) and 551 meV (HSE06 on top of PBE).

Moreover, an analysis of the curvature of the bands from the ARPES measurements also allows us to deduce the effective mass of the single layer $WS_2$ forming the heterostructure. We determined an experimentally derived hole effective mass of 0.4 $m_0$ (upper band) and 0.5 $m_0$ (lower band) (where $m_0$ is the free electron mass) at *K*, and a hole effective mass of 1.7 $m_0$ at Γ. These values agree very well with the PBE calculated bands, suggesting that many body effects are not important close to the top of the valence band at K.

Finally, the gap between the two highest energy valence bands and the bands at lower energies is substantially underestimated in PBE. The inclusion of the exchange interaction within the HSE06 functional on top of the PBE geometry corrects both errors but leads to a somewhat too large spin-orbit coupling.

Although we cannot include explicitly the interaction with the substrate as we miss the structural orientation between $WS_2$ and graphene, it seems that the measured ARPES data agrees well with the bands of an isolated single layer with PBE geometry and electronic structure obtained with the inclusion of the exchange interaction. In particular the exchange interaction seems to improve the agreement not only for the top of the valence band at zone center but also for the electronic structure in the binding energy region below -3 eV.

## IV. CONCLUSIONS

In summary, we have studied the electronic structure of single layer $WS_2$ on epitaxial graphene. We found that this heterostructure gives rise to sharp bands, in particular for the $WS_2$ VBM near the $\Gamma$ and K points. We directly observe the strong spin splitting of the upper VB and its value are in excellent agreement with the PBE DFT calculations. Our ARPES measurements on the heterostructure showed graphene and $WS_2$ largely retained their original electronic structure and, in particular suspended $WS_2$ monolayer is a good approximation of the electronic properties close to the top of the valence band. Finally, we determined experimentally spin-orbit splitting. Our results provide an important reference for future studies of electronic properties of $WS_2$ and their applications in spintronic and valleytronic devices

**Acknowledgements:** We acknowledge support from the Agence Nationale de la Recherche (ANR) under grants Labex Nanosaclay and H2DH (ANR-15-CE24-0016), from the Region Ile-de-France in the framework of C'Nano IdF (nanoscience competence center of Paris Region). Labex Nanoscalay belongs to the public funded Investissements d'Avenir program managed by ANR.

**Figure captions:**

**Figure 1: Structural and electronic properties of a WS2/graphene heterostructure:** a) crystalline structure of $WS_2$, b) typical optical image of the $WS_2$ transferred onto the graphene layer. The contrast has been adjusted in order to improve the visibility of the flake. The $WS_2$ profile is traced by white dashed lines, c) micro-Raman spectra taken on different position of the WS2/graphene heterostructure, in the top spectrum the used multi-peak Lorentzian fit is shown as blue lines, d) Raman maps images of the peak intensity and position of the 2LA(M) (top) and $A_{1g}$ ($\Gamma$) (bottom) modes of $WS_2$ on epitaxial graphene

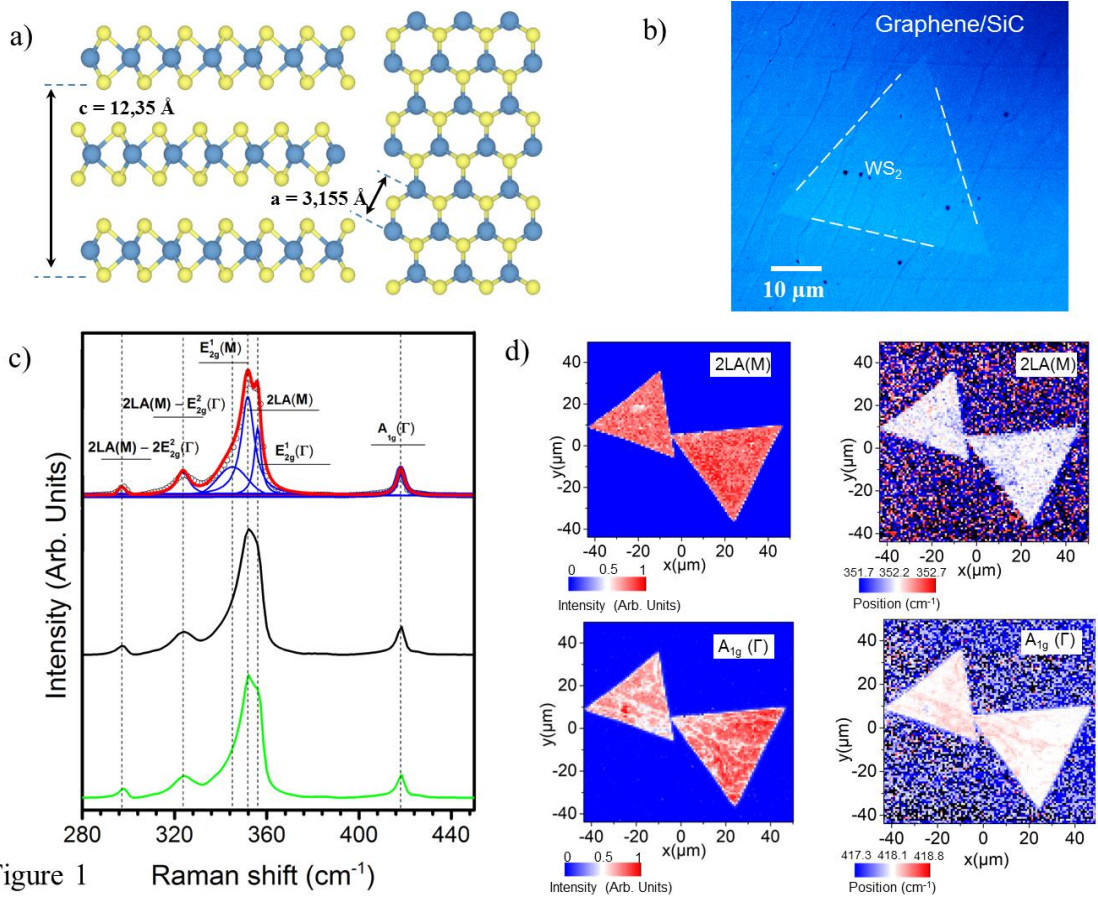

Figure 1

**Figure 2: High resolution XPS and ARPES of WS$_2$/graphene heterostructures**: a) C-1s and b) W-4f , (e) S-2p core levels at hν = 340 eV.d) ARPES map of WS$_2$/graphene heterostructure around the Γ point in the ΓK high symmetry direction of WS$_2$ Brillouin zone, e) Band structure of the graphene layer in the heterostructure at the ΓK high symmetry direction, f) corresponding momentum distribution curves (MDCs) of figure (e).

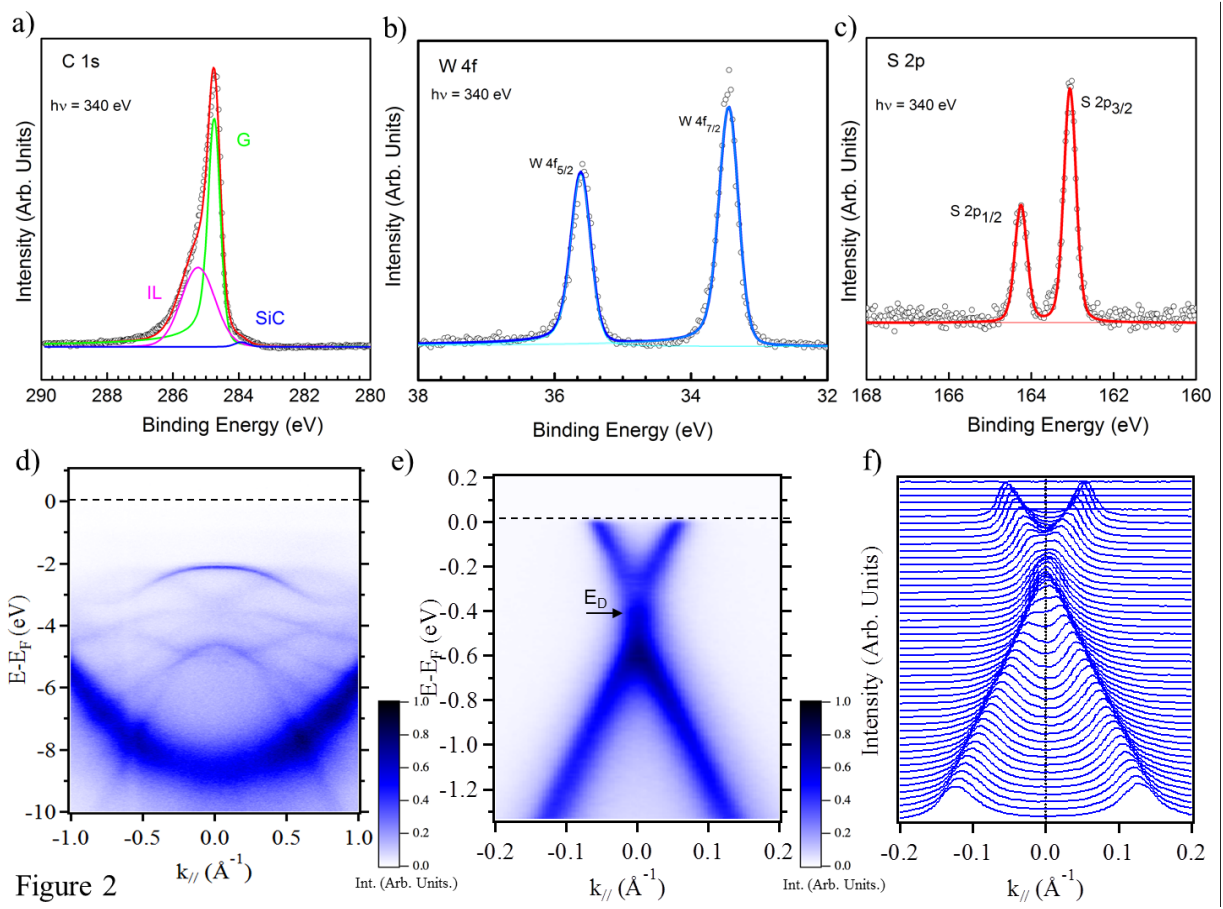

Figure 2

**Figure 3:** a) ARPES map of single layer WS$_2$ along the ΓK high symmetry directions, b) and c) Comparison between ARPES map and DFT calculations (PBE and HSE06 respectively).

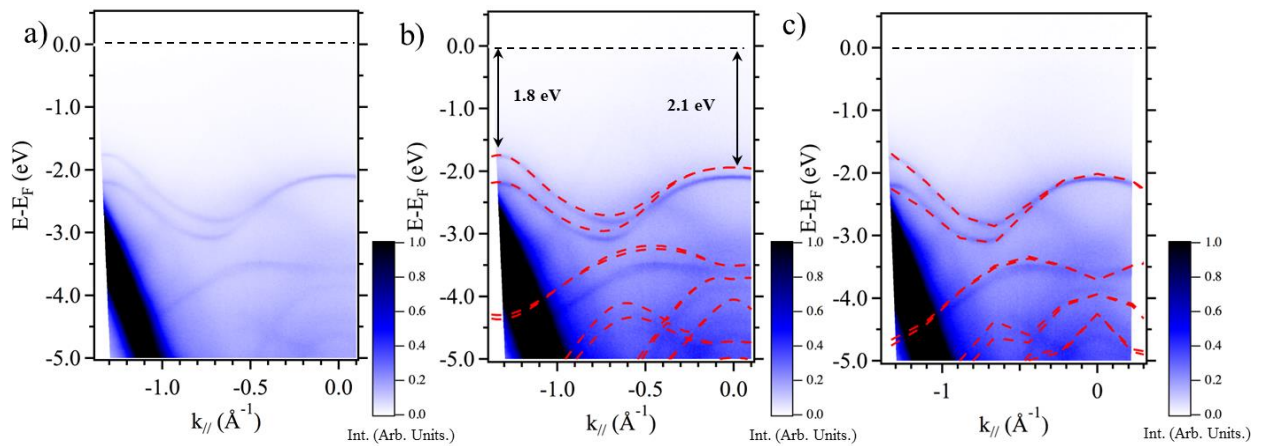

## References

[1]   A. K. Geim and I. V. Grigorieva, Van der Waals heterostructures., Nature **499**, 419 (2013).

[2]   H. Li, Q. Zhang, C. C. R. Yap, B. K. Tay, T. H. T. Edwin, A. Olivier, and D. Baillargeat, From Bulk to Monolayer MoS2: Evolution of Raman Scattering, Adv. Funct. Mater. **22**, 1385 (2012).

[3]   E. V Castro, K. S. Novoselov, S. V Morozov, N. M. R. Peres, J. M. B. L. Santos, J. Nilsson, A. K. Geim, and A. H. C. Neto, Biased Bilayer Graphene : Semiconductor with a Gap Tunable by the Electric Field Effect, Phys. Rev. Lett. **99**, 216802 (2007).

[4] L. Yang, N. a. Sinitsyn, W. Chen, J. Yuan, J. Zhang, J. Lou, and S. A. Crooker, Long-lived nanosecond spin relaxation and spin coherence of electrons in monolayer MoS2 and WS2, Nat. Phys. **11**, 830 (2015).

[5] J. He, D. He, Y. Wang, and H. Zhao, Probing effect of electric field on photocarrier transfer in graphene-WS_2 van der Waals heterostructures, Opt. Express **25**, 1949 (2017).

[6] G. Froehlicher, E. Lorchat, and S. Berciaud, Charge Versus Energy Transfer in Atomically Thin Graphene-Transition Metal Dichalcogenide van der Waals Heterostructures, Phys. Rev. X **8**, 11007 (2018).

[7] E. Pallecchi, F. Lafont, V. Cavaliere, F. Schopfer, D. Mailly, W. Poirier, and A. Ouerghi, High Electron Mobility in Epitaxial Graphene on 4H-SiC(0001) via post-growth annealing under hydrogen., Sci. Rep. **4**, 4558 (2014).

[8] T. Georgiou, R. Jalil, B. D. Belle, L. Britnell, R. V Gorbachev, S. V Morozov, Y.-J. Kim, A. Gholinia, S. J. Haigh, O. Makarovsky, et al., Vertical field-effect transistor based on graphene-WS2 heterostructures for flexible and transparent electronics, Nat. Nanotechnol. **8**, 100 (2012).

[9] Z. Hu, Q. Liu, S. Chou, and S. Dou, Advances and Challenges in Metal Sulfides / Selenides for Next-Generation Rechargeable Sodium-Ion Batteries, Adv. Mater. **1700606**, 1 (2017).

[10] S. Ulstrup, J. Katoch, R. J. Koch, D. Schwarz, S. Singh, K. M. McCreary, H. K. Yoo, J. Xu, B. T. Jonker, R. K. Kawakami, et al., Spatially Resolved Electronic Properties of Single-Layer WS2 on Transition Metal Oxides, ACS Nano **10**, 10058 (2016).

[11] J. Katoch, S. Ulstrup, R. J. Koch, S. Moser, K. M. McCreary, S. Singh, J. Xu, B. T. Jonker, R. K. Kawakami, A. Bostwick, et al., Giant spin-splitting and gap renormalization driven by trions in single-layer WS$_2$/h-BN heterostructures, arXiv 1 (2017).

[12] D. Pierucci, H. Henck, J. Avila, A. Balan, C. H. Naylor, G. Patriarche, Y. J. Dappe, M. G. Silly, F. Sirotti, A. T. C. Johnson, et al., Band alignment and minigaps in monolayer MoS2-graphene van der Waals heterostructures, Nano Lett. **16**, 4054 (2016).

[13] A. Grubišić Čabo, J. A. Miwa, S. S. Grønborg, J. M. Riley, J. C. Johannsen, C. Cacho, O. Alexander, R. T. Chapman, E. Springate, M. Grioni, et al., Observation of Ultrafast Free Carrier Dynamics in Single Layer MoS $_2$, Nano Lett. **15**, 5883 (2015).

[14] W. Jin, P. Yeh, N. Zaki, D. Chenet, G. Arefe, Y. Hao, A. Sala, T. O. Mentes, J. I. Dadap, A. Locatelli, et al., Tuning the electronic structure of monolayer graphene / MoS 2 van der Waals heterostructures via interlayer twist**201409**, 1 (2015).

[15] P. C. Yeh, W. Jin, N. Zaki, D. Zhang, J. T. Liou, J. T. Sadowski, A. Al-Mahboob, J. I. Dadap, I. P. Herman, P. Sutter, et al., Layer-dependent electronic structure of an atomically heavy two-dimensional dichalcogenide, Phys. Rev. B - Condens. Matter Mater. Phys. **91**, 1 (2015).

[16] S. Forti, A. Rossi, H. Büch, T. Cavallucci, F. Bisio, A. Sala, T. O. Menteş, A. Locatelli, M. Magnozzi, M. Canepa, et al., Electronic properties of single-layer tungsten disulfide on epitaxial graphene on silicon carbide, Nanoscale **9**, 16412 (2017).

[17] A. Rossi, D. Spirito, F. Bianco, S. Forti, F. Fabbri, H. Büch, A. Tredicucci, R. Krahne, and C. Coletti, Patterned tungsten disulfide/graphene heterostructures for efficient multifunctional optoelectronic devices, Nanoscale **10**, 4332 (2018).

[18] K. Liu, Q. Yan, M. Chen, W. Fan, Y. Sun, J. Suh, D. Fu, S. Lee, J. Zhou, S. Tongay, et al., Elastic Properties of Chemical-Vapor-Deposited Monolayer MoS2 , WS2 , and Their Bilayer Heterostructures, Nano Lett. **14**, 5097 (2014).

[19] M. T. Dau, C. Vergnaud, A. Marty, F. Rortais, C. Beigné, H. Boukari, E. Bellet-Amalric, V. Guigoz, O.

Renault, C. Alvarez, et al., Millimeter-scale layered MoSe2 grown on sapphire and evidence for negative magnetoresistance, Appl. Phys. Lett. **110**, 11909 (2017).

[20] T. L. E. Strategy, K. Chen, X. Wan, J. Wen, W. Xie, Z. Kang, X. Zeng, and C. E. T. Al, Electronic Properties of MoS2- WS2 Heterostructures Synthesized with Two-Step Lateral Epitaxial Strategy, ACS Nano **9**, 9868 (2015).

[21] X. Zhang, F. Meng, J. R. Christianson, C. Arroyo-Torres, M. a Lukowski, D. Liang, J. R. Schmidt, and S. Jin, Vertical heterostructures of layered metal chalcogenides by van der Waals epitaxy., Nano Lett. **14**, 3047 (2014).

[22] H. Sediri, D. Pierucci, M. Hajlaoui, H. Henck, G. Patriarche, Y. J. Dappe, S. Yuan, B. Toury, R. Belkhou, M. G. Silly, et al., Atomically Sharp Interface in an h-BN-epitaxial graphene van der Waals Heterostructure., Sci. Rep. **5**, 16465 (2015).

[23] E. Pallecchi, M. Ridene, D. Kazazis, C. Mathieu, F. Schopfer, W. Poirier, D. Mailly, and A. Ouerghi, Observation of the quantum Hall effect in epitaxial graphene on SiC(0001) with oxygen adsorption, Appl. Phys. Lett. **100**, 253109 (2012).

[24] D. Pierucci, H. Sediri, M. Hajlaoui, E. Velez-fort, Y. J. Dappe, M. G. Silly, R. Belkhou, A. Shukla, F. Sirotti, N. Gogneau, et al., Self-organized metal – semiconductor epitaxial graphene layer on off-axis 4H-SiC ( 0001 ), Nano Res. **8**, 1026 (2015).

[25] D. Pierucci, H. Henck, C. H. Naylor, H. Sediri, E. Lhuillier, A. Balan, J. E. Rault, Y. J. Dappe, F. Bertran, P. Le Févre, et al., Large area molybdenum disulphide - epitaxial graphene vertical Van der Waals heterostructures, Sci. Rep. **6**, 26656 (2016).

[26] See Supplemental Material at [URL will be inserted by publisher] for details on the fabrication process, measurements and calculations.

[27] G. H. Han, N. J. Kybert, C. H. Naylor, B. S. Lee, J. Ping, J. H. Park, J. Kang, S. Y. Lee, Y. H. Lee, R. Agarwal, et al., Seeded growth of highly crystalline molybdenum disulphide monolayers at controlled locations., Nat. Commun. **6**, 6128 (2015).

[28] F. Schwierz, J. Pezoldt, and R. Granzner, Two-dimensional materials and their prospects in transistor electronics, Nanoscale **7**, 8261 (2015).

[29] M. O'Brien, N. McEvoy, D. Hanlon, T. Hallam, J. N. Coleman, and G. S. Duesberg, Mapping of Low-Frequency Raman Modes in CVD-Grown Transition Metal Dichalcogenides: Layer Number, Stacking Orientation and Resonant Effects, Sci. Rep. **6**, 19476 (2016).

[30] W. Zhao, Z. Ghorannevis, K. K. Amara, J. R. Pang, M. Toh, X. Zhang, C. Kloc, P. H. Tan, and G. Eda, Lattice dynamics in mono- and few-layer sheets of WS2 and WSe2, Nanoscale **5**, 9677 (2013).

[31] Y. Kobayashi, S. Sasaki, S. Mori, H. Hibino, Z. Liu, K. Watanabe, T. Taniguchi, K. Suenaga, Y. Maniwa, and Y. Miyata, Growth and Optical Properties of High-Quality Monolayer WS 2 on Graphite, ACS Nano **9**, 4056 (2015).

[32] H. R. Gutie, N. Perea-lo, A. Laura, A. Berkdemir, B. Wang, and M. Terrones, Extraordinary Room-Temperature Photoluminescence in Triangular WS 2 Monolayers(2013).

[33] A. Molina-Sánchez and L. Wirtz, Phonons in single-layer and few-layer MoS2 and WS2, Phys. Rev. B **84**, 155413 (2011).

[34] A. Berkdemir, H. R. Gutiérrez, A. R. Botello-Méndez, N. Perea-López, A. L. Elías, C.-I. Chia, B. Wang, V. H. Crespi, F. López-Urías, J.-C. Charlier, et al., Identification of individual and few layers of WS2 using Raman spectroscopy., Sci. Rep. **3**, 1755 (2013).

[35] F. Withers, T. H. Bointon, D. C. Hudson, M. F. Craciun, and S. Russo, Electron transport of WS2 transistors in a hexagonal boron nitride dielectric environment, Sci. Rep. **4**, 4967 (2015).

[36] G. V. Bianco, M. Losurdo, M. M. Giangregorio, A. Sacchetti, P. Prete, N. Lovergine, P. Capezzuto, and G. Bruno, Direct epitaxial CVD synthesis of tungsten disulfide on epitaxial and CVD graphene, RSC Adv. **5**, 98700 (2015).

[37] F. Withers, T. H. Bointon, D. C. Hudson, M. F. Craciun, and S. Russo, Electron transport of WS2 transistors in a hexagonal boron nitride dielectric environment, Sci. Rep. **4**, 4967 (2015).

[38] X. Mao, Y. Xu, Q. Xue, W. Wang, and D. Gao, Ferromagnetism in exfoliated tungsten disulfide nanosheets, Nanoscale Res. Lett. **8**, 430 (2013).

[39] Y. Li, D. Chen, and R. A. Caruso, Enhanced electrochromic performance of WO3 nanowire networks grown directly on fluorine-doped tin oxide substrates, J. Mater. Chem. C **4**, 10500 (2016).

[40] D. Barrera, Q. Wang, Y.-J. Lee, L. Cheng, M. J. Kim, J. Kim, and J. W. P. Hsu, Solution synthesis of few-layer 2H MX2 (M = Mo, W; X = S, Se), J. Mater. Chem. C **5**, 2859 (2017).

[41] Y. Gao, Z. Liu, D.-M. Sun, L. Huang, L.-P. Ma, L.-C. Yin, T. Ma, Z. Zhang, X.-L. Ma, L.-M. Peng, et al., Large-area synthesis of high-quality and uniform monolayer WS2 on reusable Au foils **6**, 8569 (2015).

[42] S. Y. Zhou, G.-H. Gweon, A. V. Fedorov, P. N. First, W. A. De Heer, D.-H. Lee, F. Guinea, A. H. C. Neto, and A. Lanzara, Substrate-induced bandgap opening in epitaxial graphene., Nat. Mater. **6**, 770 (2007).

[43] H. Coy Diaz, R. Addou, and M. Batzill, Interface properties of CVD grown graphene transferred onto MoS2(0001)., Nanoscale **6**, 1071 (2014).

[44] H. C. Diaz, J. Avila, C. Chen, R. Addou, M. C. Asensio, and M. Batzill, Direct observation of interlayer hybridization and Dirac relativistic carriers in graphene/MoS$_2$ van der Waals heterostructures., Nano Lett. **15**, 1135 (2015).

[45] P. Giannozzi, S. Baroni, N. Bonini, M. Calandra, R. Car, C. Cavazzoni, D. Ceresoli, G. L. Chiarotti, M. Cococcioni, I. Dabo, et al., QUANTUM ESPRESSO: a modular and open-source software project for quantum simulations of materials., J. Phys. Condens. Matter **21**, 395502 (2009).

[46] O. Andreussi, T. Brumme, O. Bunau, M. Buongiorno Nardelli, M. Calandra, R. Car, C. Cavazzoni, D. Ceresoli, M. Cococcioni, N. Colonna, et al., Advanced capabilities for materials modelling with Quantum ESPRESSO, J. Phys. Condens. Matter (2017).

[47] J. P. Perdew, K. Burke, and M. Ernzerhof, Generalized Gradient Approximation Made Simple [Phys. Rev. Lett. 77, 3865 (1996)], Phys. Rev. Lett. **78**, 1396 (1997).

[48] A. V. Krukau, O. A. Vydrov, A. F. Izmaylov, and G. E. Scuseria, Influence of the exchange screening parameter on the performance of screened hybrid functionals, J. Chem. Phys. **125**, 224106 (2006).

[49] L. Lin, Adaptively Compressed Exchange Operator, J. Chem. Theory Comput. **12**, 2242 (2016).

[50] A. Raja, A. Chaves, J. Yu, G. Arefe, H. M. Hill, A. F. Rigosi, T. C. Berkelbach, P. Nagler, C. Schüller, T. Korn, et al., Coulomb engineering of the bandgap and excitons in two-dimensional materials, Nat. Commun. **8**, 15251 (2017).

[51] S. Ulstrup, A. G. Čabo, D. Biswas, J. M. Riley, M. Dendzik, C. E. Sanders, M. Bianchi, C. Cacho, D. Matselyukh, R. T. Chapman, et al., Spin and valley control of free carriers in single-layer WS2, Phys. Rev. B **95**, 1 (2017).

# Supplementary Information

# Electronic band structure of Two-Dimensional WS$_2$/Graphene van der Waals Heterostructures

Hugo Henck[1], Zeineb Ben Aziza[1], Debora Pierucci[2], Feriel Laourine[1], Francesco Reale[3] and Pawel Palczynski[3], Julien Chaste[1], Mathieu G. Silly[4], François Bertran[4], Patrick Le Fevre[4], Emmanuel Lhuillier[5], Taro Wakamura[6], Cecilia Mattevi[3], Julien E. Rault[4], Matteo Calandra[5] and Abdelkarim Ouerghi[1,*]

[1]Centre de Nanosciences et de Nanotechnologies, CNRS, Univ. Paris-Sud, Université Paris-Saclay, C2N – Marcoussis, 91460 Marcoussis, France
[2]CELLS - ALBA Synchrotron Radiation Facility, Carrer de la Llum 2-26, 08290 Cerdanyola del Valles, Barcelona, Spain
[3]Imperial College London Department of Materials, Exhibition road London SW7 2AZ, UK
[4]Synchrotron-SOLEIL, Saint-Aubin, BP48, F91192 Gif sur Yvette Cedex, France
[5]Sorbonne Universités, UPMC Univ. Paris 06, CNRS-UMR 7588, Institut des NanoSciences de Paris, 4 place Jussieu, 75005 Paris, France
[6]Laboratoire de Physique des Solides, CNRS, Université Paris-Sud, Université Paris Saclay, 91405 Orsay Cedex, France

*Corresponding author, E-mail: abdelkarim.ouerghi@c2n.upsaclay.fr,

## I.  Transfer process of WS$_2$ on epitaxial graphene:

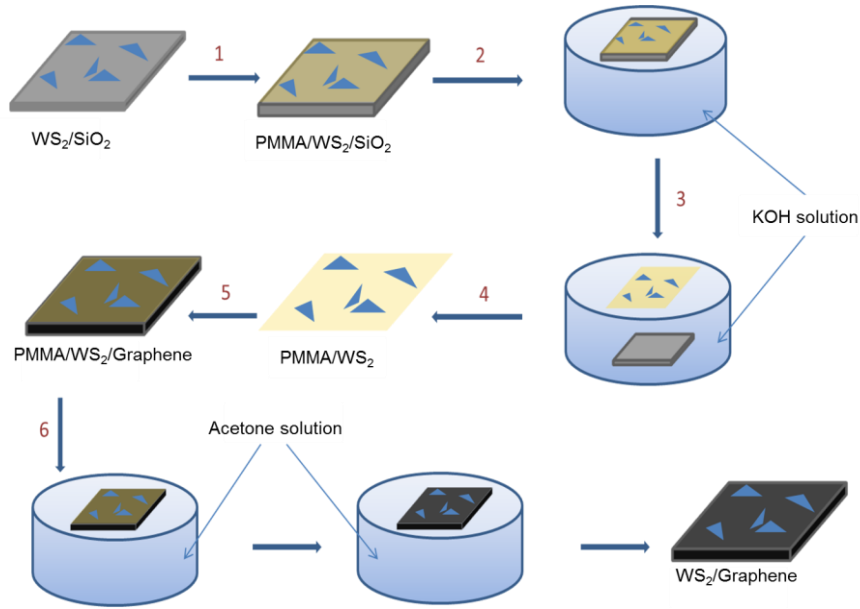

**Figure S1**: Schematic transfer process of WS$_2$ on epitaxial graphene

For the transfer process PMMA was spin-coated onto the WS$_2$ flakes (step 1) and peeled them off from the SiO$_2$ substrate by wet etching in KOH solution (step 2 and 3). Afterward, the PMMA/WS$_2$ layer was transferred onto the graphene/SiC substrate (step 4 and 5). The PMMA was finally removed using acetone (Step 6).

## II. Micro-Raman spectroscopy

The micro-Raman spectra measurements were performed on the Renishaw microscope using a 532 nm laser in an ambient environment at room temperature. The excitation laser was focused onto the samples with spot diameter of ~1 μm and incident power of ~400 μW. The integration time was optimized to obtain a satisfactory signal-to-noise ratio. We obtained Raman spatial maps by raster scanning with 0.3μm step size using a precision 2D mapping stage.

Using Raman spectroscopy, in the case of 2D material, it is possible to discriminate the strain, doping and temperature variation. For this purpose, we have plotted the map of the intensity and position of the three Raman active mode of WS$_2$ ($E^1_{2g}(\Gamma)$, $2LA(M)$, $A_{1g}(\Gamma)$) in figure 2 (d) and S2.

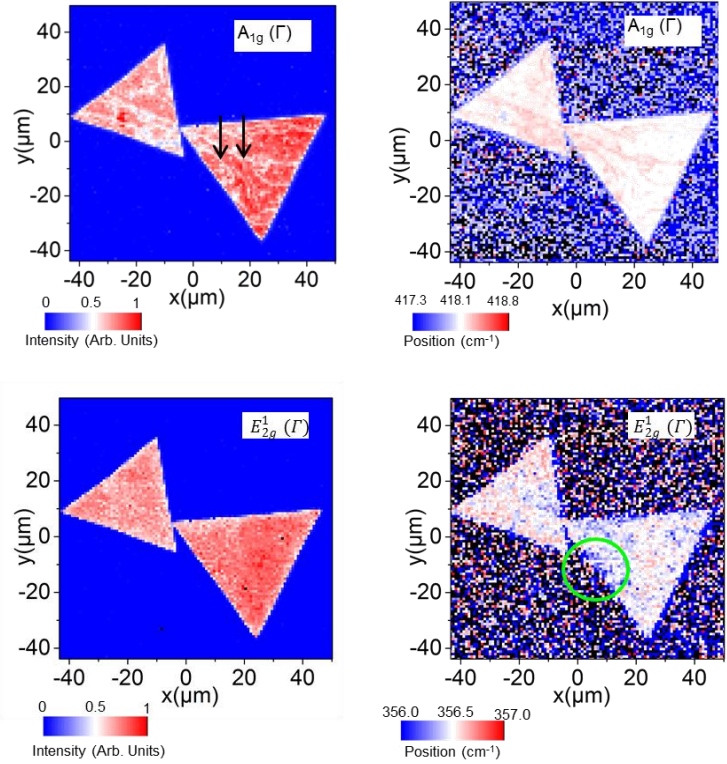

**Figure S2:** Raman maps images of the peak intensity and position of the $A_{1g}(\Gamma)$ (top) and $E^1_{2g}(\Gamma)$ (bottom) modes of WS$_2$ on epitaxial graphene.

The WS$_2$ flakes on graphene are clearly homogenous, very clean, without significant properties variation. To go beyond this structural characterization we analyzed the relative Raman peak shift. This procedure are widely used for graphene [1] and can be extended to other 2D material as MoS$_2$ and WS$_2$. The relative variation between the different Raman peaks for a monolayer WS$_2$ depends strongly on the cause of this shift. If we consider the ratio between $\Delta A_{1g}(\Gamma)/\Delta E^1_{2g}(\Gamma) = \Delta$, this value changes when the WS$_2$ monolayer is subject to strain, temperature or doping variation as shown in previous studies:

- Temperature variation [2,3] : $\Delta$= 1.33 and 1.3
- Doping variation under UV [4] : $\Delta$= 2.23
- Strain variation [5,6] : $\Delta$= 0.29 and ~ 0.3

The shift in position of the peak $2LA\ (M)$ and $E^1_{2g}\ (\Gamma)$ is almost the same. We plot in figure S3 the position of the $A_{1g}\ (\Gamma)$ as a function of the position of the $E^1_{2g}\ (\Gamma)$ peak. Three different slope of the date can be identified in the graph. The blue line indicates the expected variation due to the strain, the red line due to temperature variation and the green one to doping. A clear signature of strain and charge doping can be identified in some areas of our sample. The main strained region is pointed in Figure S2 by a green circle and the doped region is related to the SiC step position along the substrate (indicated by black arrows in Figure S2). If we excluded these specific regions, the $WS_2$ flakes are very clean and homogenous.

If we take as a reference the strain-shift relation to be -2.07cm$^{-1}$/% of strain for the $E^1_{2g}$ peak [5,6], we can quantify the strain to be ±0.05% (corresponding to a $\Delta$=±0.1cm$^{-1}$) on almost all the sample. In the maximum strained region (blue region), the value is around 0.6% ($\Delta$=1.3cm$^{-1}$).

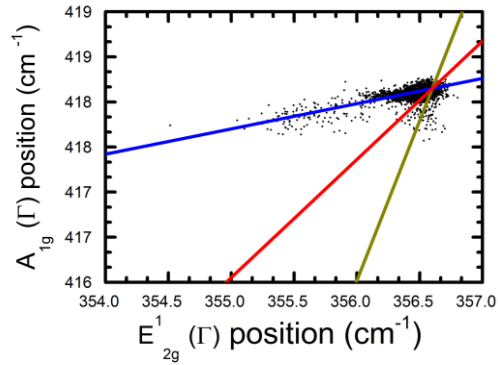

**Figure S3:** Lee et al. diagram [1] related to the two $WS_2$ flakes showed in Figure S2. $A_{1g}\ (\Gamma)$ peak position as a function of the $E^1_{2g}\ (\Gamma)$ peak position (black dots) and the expected variation for strain (blue line), doping (green line) and temperature (red line).

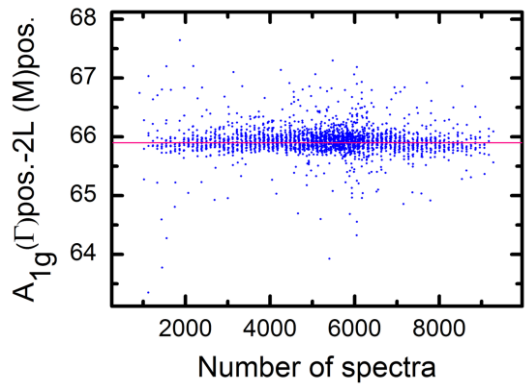

**Figure S4:** Calculated frequency difference between the $A_{1g}\ (\Gamma)$ and the $2LA\ (M)$ mode referring to all the spectra collected in the Raman map of figure 2(d). An average value of 65.9 ± 0.1 cm$^{-1}$ is obtained.

## III.  Optical band gap:

The PL measurements were carried out using a confocal commercial Renishaw micro-Raman microscope with a 100× objective and a Si detector

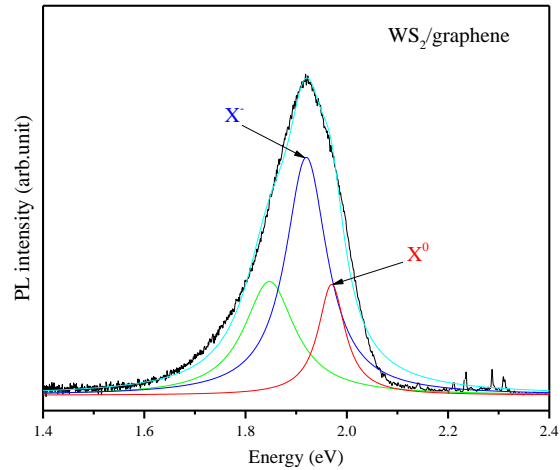

**Figure S5:** Photoluminescence spectra of $WS_2$/graphene and $WS_2$/$SiO_2$.

From Figure S5, we can see that the PL intensity of $WS_2$ on graphene (red line) shows a drastic quench with respect to $WS_2$/$SiO_2$ (blue line). This can be attributed an interlayer interaction between $WS_2$ and n doped graphene substrate which leads to an electron transfer from graphene to $WS_2$. [7,8] To investigate this eventual charge transfer, we performed fits on $WS_2$ PL as shown in the SI in Figure S6. It is clearly obvious that when using graphene as a substrate, the emission of exciton ratio ($X^0$)/ trion (X) [9,10] increases. This observation confirms that electron transfer happened from n doped graphene to $WS_2$ [11]. Moreover, since the PL energy and intensity vary between the two different substrates, we cannot completely exclude the possibility of stress level variation sustained by $WS_2$ flakes. [12]

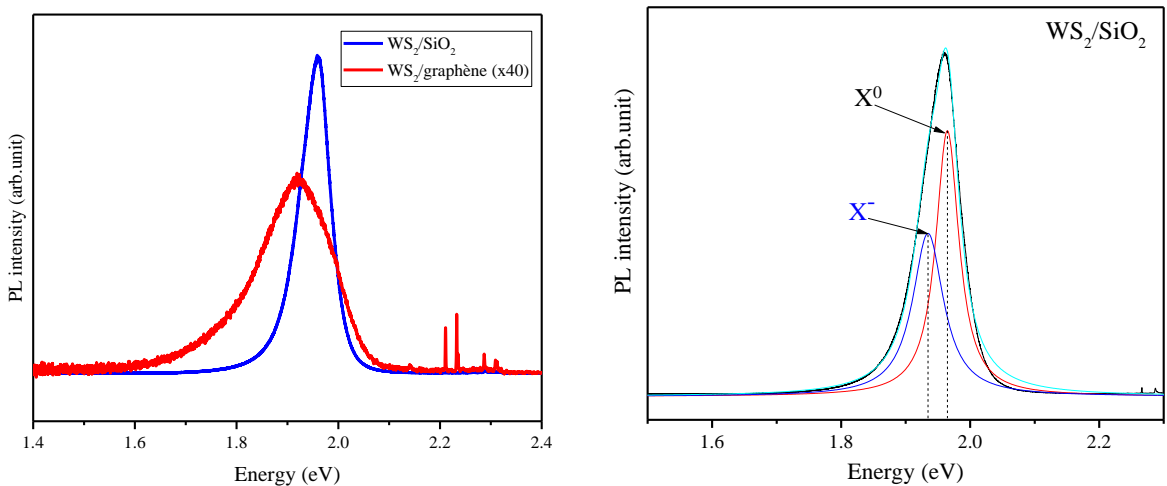

**Figure S6:** Fits of photoluminescence spectra of $WS_2$/graphene and $WS_2$/$SiO_2$.

## IV. <u>EDC curve at the Γ point of the $WS_2$ Brillouin zone:</u>

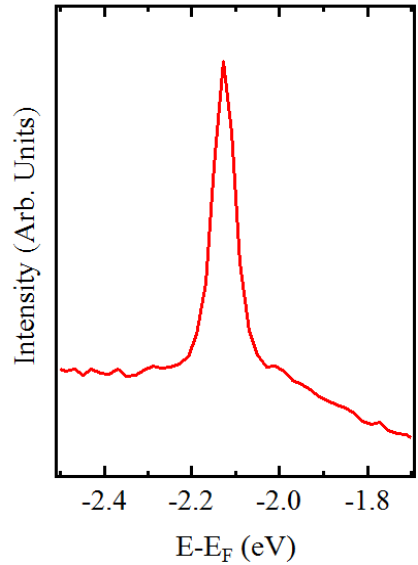

**Figure S7:** Energy distribution curves (EDC) extracted from the ARPES map (Figure 2 (b)) at the Γ point.

## V.  Work Function measurement:

The work function of the $WS_2$/graphene was determined *via* the measurement of the low energy cut-off of the secondary electron (SE) energy distribution curve obtained by XPS measurements (Figure S7). To be sure that the SE have a kinetic energy higher than the spectrometer vacuum level, the sample is negatively biased (-10 V) with respect to the analyzer. A value of 4.1 ± 0.1 eV is found, which is similar to the value of the pristine n-doped graphene [13].

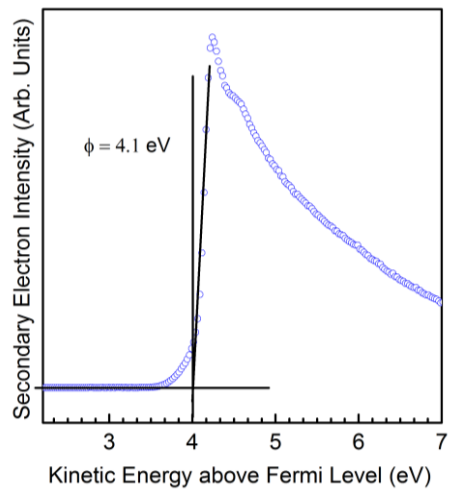

**Figure S8:** Secondary electron cut-off for determining the work function of the $WS_2$/graphene heterostructure.

## VI.  Electronic structure calculations: technical details.

Electronic structure calculations were performed using the QUANTUM-ESPRESSO code [14,15] and the PBE [16] and HSE06 [17] functional. Structural parameters are the same as in Ref. [18]. We use ONCV norm-conserving pseudopotentials [19–21], 18x18 k-point sampling and a 50 Ryd. Cutoff. In the hybrid functionals calculations, we use the adaptive compressed exchange [22]. In evaluating the exchange interaction we perform downsampling as we used a 100 Ryd cutoff (instead of 200 Ryd) and 9x9 a k+q point grid.

**References:**

[1] J. E. Lee, G. Ahn, J. Shim, Y. S. Lee, and S. Ryu, Optical separation of mechanical strain from charge doping in graphene, Nat. Commun. **3**, 1024 (2012).

[2] X. Huang, Y. Gao, T. Yang, W. Ren, H.-M. Cheng, and T. Lai, Quantitative Analysis of Temperature Dependence of Raman shift of monolayer $WS_2$, Sci. Rep. **6**, 32236 (2016).

[3] Z. Hu, Y. Bao, Z. Li, Y. Gong, R. Feng, Y. Xiao, X. Wu, Z. Zhang, X. Zhu, P. M. Ajayan, et al., Temperature dependent Raman and photoluminescence of vertical WS2/MoS2 monolayer heterostructures, Sci. Bull. **62**, 16 (2017).

[4] M. W. Iqbal, M. Z. Iqbal, M. F. Khan, M. A. Shehzad, Y. Seo, and J. Eom, Deep-ultraviolet-light-driven reversible doping of WS2 field-effect transistors, Nanoscale **7**, 747 (2014).

[5] F. Wang, I. A. Kinloch, D. Wolverson, R. Tenne, A. Zak, E. O'Connell, U. Bangert, and R. J. Young, Strain-induced phonon shifts in tungsten disulfide nanoplatelets and nanotubes, 2D Mater. **4**, 1 (2016).

[6] Y. Wang, C. Cong, W. Yang, J. Shang, N. Peimyoo, Y. Chen, J. Kang, J. Wang, W. Huang, and T. Yu, Strain-induced direct–indirect bandgap transition and phonon modulation in monolayer $WS_2$, Nano Res. **8**, 2562 (2015).

[7] P. Hu, J. Ye, X. He, K. Du, K. K. Zhang, X. Wang, Q. Xiong, Z. Liu, H. Jiang, and C. Kloc, Control of Radiative Exciton Recombination by Charge Transfer Induced Surface Dipoles in MoS2 and WS2 Monolayers, Sci. Rep. **6**, 24105 (2016).

[8] H. Tan, Y. Fan, Y. Rong, B. Porter, C. S. Lau, Y. Zhou, Z. He, S. Wang, H. Bhaskaran, and J. H. Warner, Doping Graphene Transistors Using Vertical Stacked Monolayer WS 2 Heterostructures Grown by Chemical Vapor Deposition, ACS Appl. Mater. Interfaces **8**, 1644 (2016).

[9] M. S. Kim, S. J. Yun, Y. Lee, C. Seo, G. H. Han, K. K. Kim, Y. H. Lee, and J. Kim, Biexciton Emission from Edges and Grain Boundaries of Triangular WS2 Monolayers, ACS Nano **10**, 2399 (2016).

[10] I. Kylänpää and H. P. Komsa, Binding energies of exciton complexes in transition metal dichalcogenide monolayers and effect of dielectric environment, Phys. Rev. B - Condens. Matter Mater. Phys. **92**, 1 (2015).

[11] J. He, N. Kumar, M. Z. Bellus, H.-Y. Chiu, D. He, Y. Wang, and H. Zhao, Electron transfer and coupling in graphene-tungsten disulfide van der Waals heterostructures., Nat. Commun. **5**, 5622 (2014).

[12] K. Kang, K. Godin, Y. D. Kim, S. Fu, W. Cha, J. Hone, and E. H. Yang, Graphene-Assisted Antioxidation of Tungsten Disulfide Monolayers: Substrate and Electric-Field Effect, Adv. Mater. **29**, (2017).

[13] E. Pallecchi, F. Lafont, V. Cavaliere, F. Schopfer, D. Mailly, W. Poirier, and A. Ouerghi, High Electron

Mobility in Epitaxial Graphene on 4H-SiC(0001) via post-growth annealing under hydrogen., Sci. Rep. **4**, 4558 (2014).

[14] P. Giannozzi, S. Baroni, N. Bonini, M. Calandra, R. Car, C. Cavazzoni, D. Ceresoli, G. L. Chiarotti, M. Cococcioni, I. Dabo, et al., QUANTUM ESPRESSO: a modular and open-source software project for quantum simulations of materials., J. Phys. Condens. Matter **21**, 395502 (2009).

[15] O. Andreussi, T. Brumme, O. Bunau, M. Buongiorno Nardelli, M. Calandra, R. Car, C. Cavazzoni, D. Ceresoli, M. Cococcioni, N. Colonna, et al., Advanced capabilities for materials modelling with Quantum ESPRESSO, J. Phys. Condens. Matter (2017).

[16] J. P. Perdew, K. Burke, and M. Ernzerhof, Generalized Gradient Approximation Made Simple [Phys. Rev. Lett. 77, 3865 (1996)], Phys. Rev. Lett. **78**, 1396 (1997).

[17] A. V. Krukau, O. A. Vydrov, A. F. Izmaylov, and G. E. Scuseria, Influence of the exchange screening parameter on the performance of screened hybrid functionals, J. Chem. Phys. **125**, 224106 (2006).

[18] T. Brumme, M. Calandra, and F. Mauri, First-principles theory of field-effect doping in transition-metal dichalcogenides: Structural properties, electronic structure, Hall coefficient, and electrical conductivity, Phys. Rev. B **91**, 155436 (2015).

[19] D. R. Hamann, Optimized norm-conserving Vanderbilt pseudopotentials, Phys. Rev. B **88**, 85117 (2013).

[20] Optimization algorithm for the generation of ONCV pseudopotentials, Comput. Phys. Commun. **196**, 36 (2015).

[21] P. Scherpelz, M. Govoni, I. Hamada, and G. Galli, Implementation and Validation of Fully Relativistic *GW* Calculations: Spin–Orbit Coupling in Molecules, Nanocrystals, and Solids, J. Chem. Theory Comput. **12**, 3523 (2016).

[22] L. Lin, Adaptively Compressed Exchange Operator, J. Chem. Theory Comput. **12**, 2242 (2016).